\documentclass[10pt,aps,prd,twocolumn,fleqn,floatfix]{revtex4-2}

\usepackage{graphicx}
\usepackage{amsmath,amssymb,amsfonts}
\usepackage{color}

\begin{document}

\title{Transition from weak to strong coupling in thermal gauge theories:\\ Lessons from ${\cal N}=4$ SYM Theory}

\author{Berndt M\"uller}
\affiliation{Department of Physics, Duke University, Durham, NC 27708}

\date{\today}

\begin{abstract}
We investigate the transition between weak and strong coupling in thermal ${\cal N}=4$ supersymmetric Yang-Mills (SYM) theory as a function of 't Hooft coupling $\lambda$ for several quantities of phenomenological interest for which next-to-leading order calculations are available in both regimes. We use Pad\'e approximants to interpolate between the weak and strong coupling expansions and determine the location and width of the transition between the asymptotic regimes.
\end{abstract}

\maketitle

The ${\cal N}=4$ supersymmetric Yang-Mills (SYM) theory has received much interest in recent decades, in part because of the viability of exact perturbative calculations for weak and strong 't Hooft coupling $\lambda = g^2 N_c$, where  $N_c$ denotes the (large) number of colors. The weak coupling expansion is the usual perturbative expansion with hard-thermal loop (HTL) modifications for finite temperature. The strong coupling expansion makes us of the Anti-de Sitter/conformal field theory (AdS/CFT) duality \cite{Maldacena:1997re,Aharony:1999ti}, in which expectation values of operators in the conformal field theory are identified with geometric objects in AdS space. At finite temperature the AdS geometry contains a large black hole whose Hawking temperature is identified with the temperature of the field theory. 

The strong coupling limit of the CFT corresponds to the classical AdS-Schwarzschild (AdS-SS) space; finite coupling corrections can be calculated as stringy corrections to the classical geometry. For quantities in the field theory whose duals can be expressed solely in terms of properties of the AdS-SS geometry, such as the equation of state or the shear viscosity, the finite coupling corrections can be obtained from quantum corrections to the Einstein-Hilbert action \cite{Gubser:1998nz,Buchel:2004di,Waeber:2015oka}. For other quantities that are dual to objects imbedded into the AdS-SS geometry, such as strings, finite coupling corrections also arise from quantum fluctuations of their world-sheets \cite{Zhang:2012jd}.

The ${\cal N}=4$ SYM theory holds valuable lessons for the border between the validity of weak and strong coupling approaches to a gauge plasma, because NLO corrections are known for many quantities of interest in both approaches. As we will see below, the transition is generally not a sharp but a gradual one -- there is no phase transition as a function of 't Hooft coupling. The central ``pseudocritical'' coupling $\lambda_c$ (to be defined below) has somewhat different values for different quantities of interest, but their transition regions between the asymptotic domains overlap.

Our strategy for mapping the border between weak and strong coupling is simple. We use the NLO weak and strong coupling results to fit the coefficients of powers of $\lambda^{1/2}$ in a generalized Pad\'e approximant that smoothly interpolates between the two limits. Our approach follows \cite{Andersen:2021bgw} where the interpolating Pad\'e approximant is constructed for the normalized entropy density $s(T,\lambda)/s(T,0) \equiv s(\lambda)/s_0$ (see Fig.~9 and Appendix C of \cite{Andersen:2021bgw}). Depending on the quantity under consideration we will then plot the first or second derivative of the Pad\'e approximant with respect to ($\ln\lambda$) and locate its extremum $\lambda_c$. It is clear that both choices, the choice of Pad\'e approximation to interpolate between weak and strong coupling and the choice of extremal property, are somewhat arbitrary. However, since the transition region has a considerable width, any reasonable choice is likely to result in values of $\lambda_c$ that fall into the same region. It is also clear that the precise values of $\lambda_c$ may change when higher-order terms in the expansions are calculated. However, we expect those changes to be well within the half-width of the peak in the function we use to locate $\lambda_c$.

The dimensionless quantities we are considering here are the normalized entropy density $s(\lambda)/s_0$, the kinematic shear viscosity $\eta/s$, the normalized diffusion constant $2\pi T D_s(T)$, and the normalized transverse broadening rate $\hat{q}/T^3$. For all these quantities, first ``stringy'' corrections of order $\lambda^{-3/2}$ from finite coupling corrections to the bulk geometry have been computed \cite{Gubser:1998nz,Buchel:2004di,Casalderrey-Solana:2006fio,Armesto:2006zv}. For $\hat{q}$ also corrections arising from world-sheet fluctuations at finite coupling  of order $\lambda^{-1/2}$ have been derived \cite{Zhang:2012jd}. Such corrections do not arise for the equation of state and the kinematic shear viscosity as these are encoded in the bulk geometry but are expected to arise for the diffusion coefficient $D_s$, for which they have not yet been calculated.

Higher-order corrections have been computed for the entropy density $s(T,\lambda)$ at NNLO \cite{Andersen:2021bgw} and at NLO for the diffusion coefficient $D_s$ \cite{Caron-Huot:2008dyw} and the jet quenching parameter $\hat{q}$ \cite{Ghiglieri:2018dib}. Only next-to-leading logarithmic results at leading order are known for the kinematic shear viscosity $\eta/s$ \cite{Huot:2006ys}, which become unphysical at rather small 't Hooft coupling and thus are insufficient for our purpose. As an alternative we make use of the scaling relation for $(\eta/s)(\hat{q}/T^3)$, which states that this combination is nearly independent of the coupling strength at weak coupling \cite{Majumder:2007zh, Muller:2021wri}. Assuming that the scaling relation also applies to the weakly coupled ${\cal N}=4$ SYM theory, we will simply use the value of this double ratio for the LO leading-log results and the known NLO expression for $hat{q}$ to set $(\eta/s)_{\rm NLO} \approx 6.173/(2\pi)(T^3/\hat{q})_{\rm NLO}$. Explicit expressions for the Pad\'e approximants are given in the Appendix. 

\begin{figure}[htb]
\centering
	\includegraphics[width=0.45\linewidth]{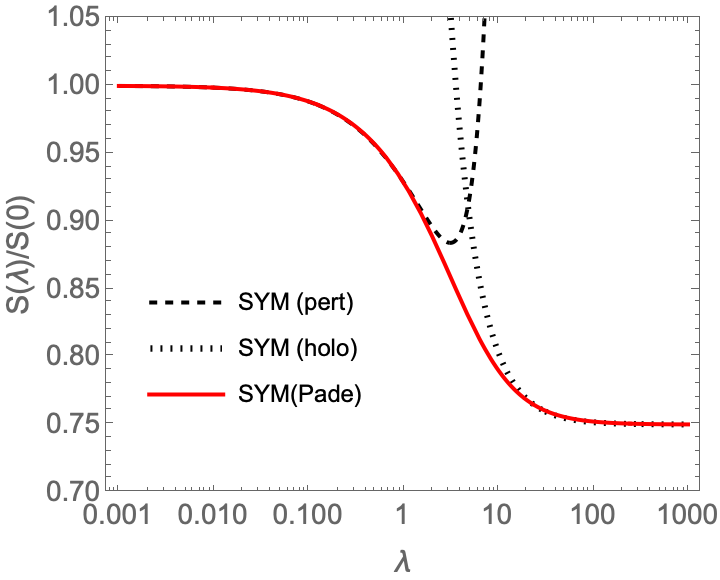}
	\hspace{0.03\linewidth}
	\includegraphics[width=0.45\linewidth]{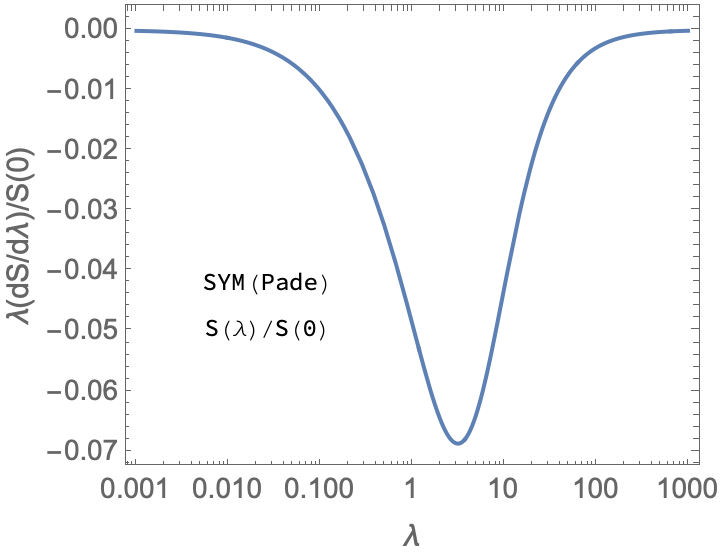}
	\hspace{0.03\linewidth}
	\includegraphics[width=0.45\linewidth]{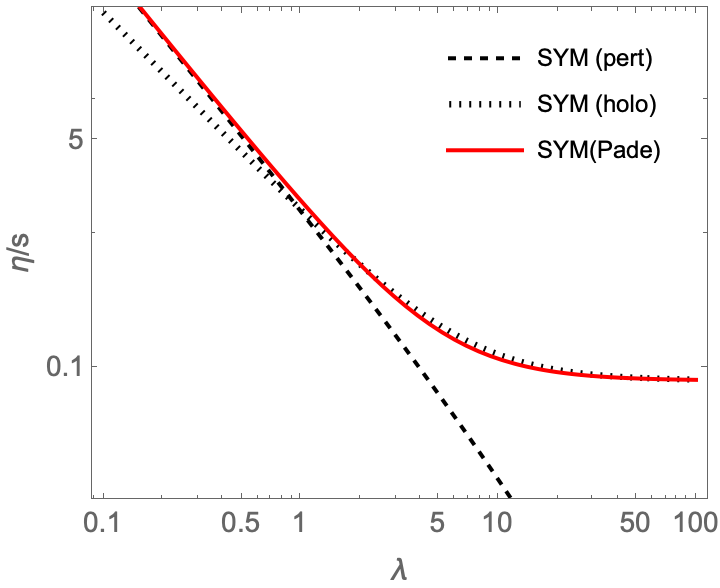}
	\hspace{0.03\linewidth}
	\includegraphics[width=0.45\linewidth]{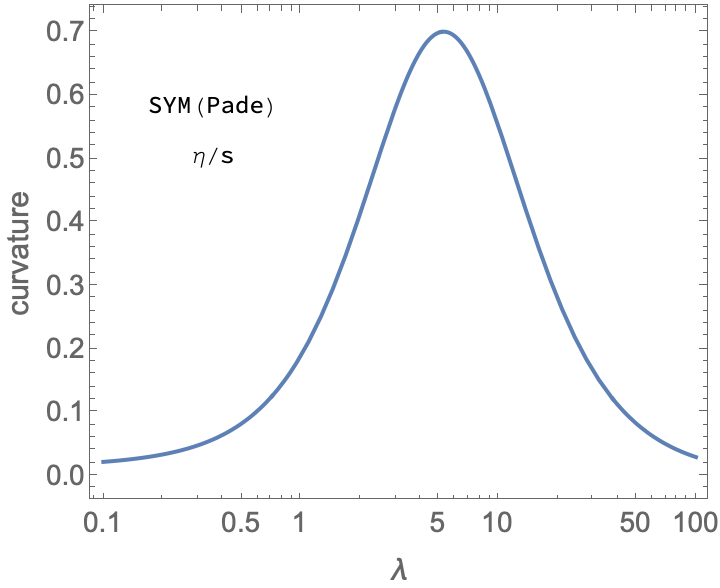}
	\hspace{0.03\linewidth}
	\includegraphics[width=0.45\linewidth]{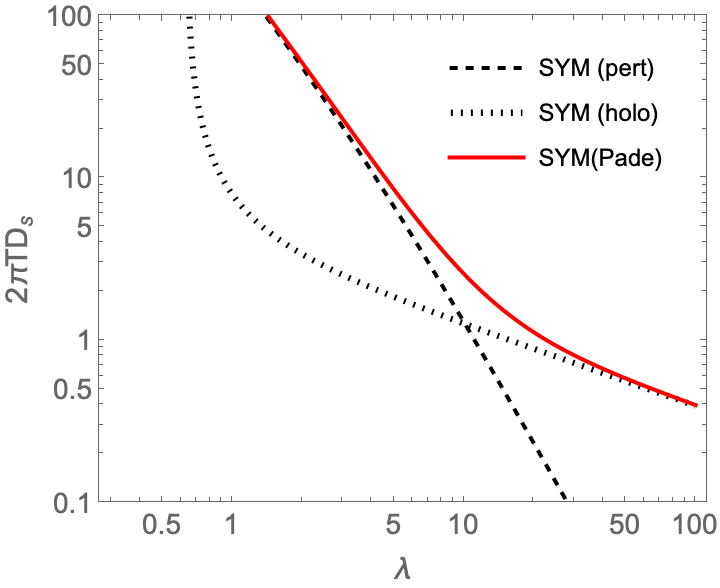}
	\hspace{0.03\linewidth}
	\includegraphics[width=0.45\linewidth]{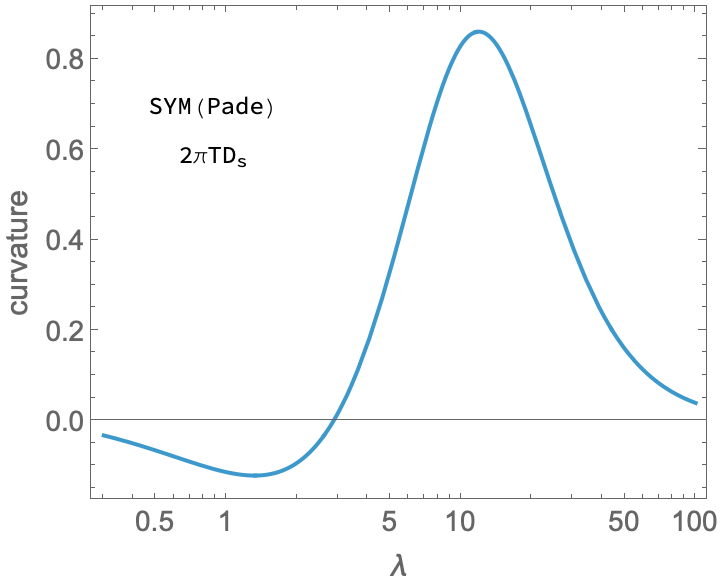}
	\hspace{0.03\linewidth}
    \includegraphics[width=0.45\linewidth]{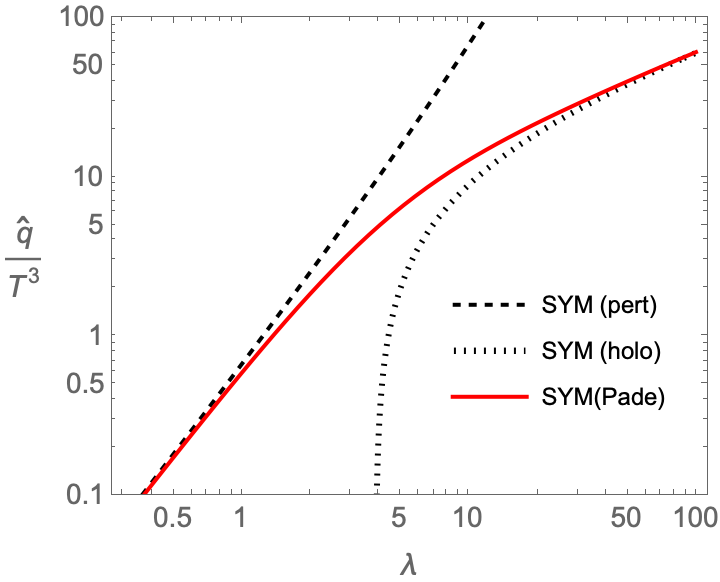}
	\hspace{0.03\linewidth}
	\includegraphics[width=0.45\linewidth]{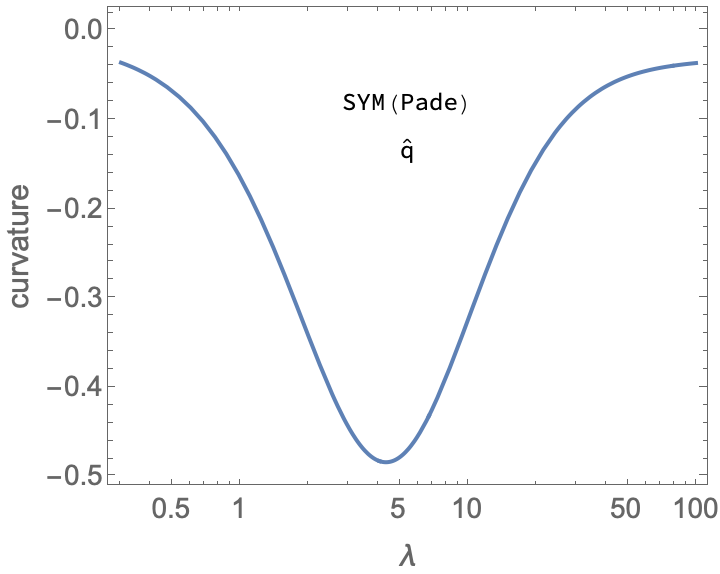}
\caption{The left column of panels shows the weak coupling (dashed curves) and strong coupling (dotted curves) expansions for (from top to bottom) $s/s_)$, $\eta/s$, $2\pi TD_s$ and $\hat{q}/T^3$ as functions of $\lambda$. The interpolating Pad\'e approximants are shown as solid red curves. The right column of panels shown the logarithmic curvature of the Pad\'e interpolations, except for $s/s_0$, where the logarithmic slope is shown. The extrema of the curves define the border $\lambda_c$ between weak and strong coupling. Numerical values for $\lambda_s$ are shown in Table~\ref{Table1}.}
\label{fig:Pade}
\end{figure}

In Fig.~\ref{fig:Pade} we show, on the left, the weak and strong coupling expansions for the quantities mentioned above as functions of $\lambda$ on a logarithmic scale. Results from the weak coupling expansion are shown as dashed lines, those from the strong coupling expansion as dotted lines, and the interpolating Pad\'e approximant is shown as solid red curve. The central transition point between the weak and strong coupling regimes for the case of the normalized entropy density $s/s_0$ (top row) is characterized by the inflection point of the red curve; we can locate it by the extremum of the first derivative of $s/s_0$ with respect to $\ln\lambda$, shown in the top right panel. We call this value $\lambda_c$. For the three transport coefficients (rows 2--4 of Fig.\ref{fig:Pade}) the transition can be characterized by an extremum of the curvature of the red curve which can be located as the extremum of the second derivative of the logarithm of the coefficient with respect to $\ln\lambda$. The numerical values of $\lambda_c$ are listed in Table~\ref{Table1} together with the lower and upper bounds $\lambda_\mp$ of the region defined as the values of $\lambda$ for which the first (second) derivative has decreased by half from the extremum.

We first note that all values for $\lambda_c$ lie in the range $3 < \lambda_c < 8$, which is not unexpected. For the transport coefficients, the lower and upper bounds differ from the central value by a factor 2--3; the transition region as defined here is slightly wider for the entropy density, but this may be an artifact caused by the limited range of this quantity. Averaging over the four quantities and including the widths of the transition regions, we obtain a general estimate for the weak-to-strong coupling transition region $3 \lesssim \lambda \lesssim 14$.

\begin{table}[htb]
\centering
\begin{tabular}{|c|c|c|c|c|}
\hline
Quantity & $\lambda_c$ & $\lambda_-$ & $\lambda_+$ & $f(\lambda_c)$ \\
\hline
$s(\lambda)/s_0$ & 3.14 & 0.54 & 12.95 & 0.859 \\
$\eta/s$ & 5.28 & 1.69 & 16.38 & 0.180 \\
$2\pi TD_s$ & 11.90 & 5.62 & 27.81 & 2.03 \\
$\hat{q}/T^3$ & 4.36 & 1.39 & 13.25 & 5.47 \\
\hline
\end{tabular}
\caption{The Table lists the central 't Hooft coupling $\lambda_c$, lower and upper transition region bounds $\lambda_\mp$, and Pad\'e interpolated value at $\lambda_c$ for the quantities considered here: normalized entropy density $s/s_0$, kinematic shear viscosity $\eta/s$, normalized diffusion coefficient $2\pi TD_s$, and normalized momentum broadening coefficient (jet quenching parameter) $\hat{q}/T^3$. The values in the Table for $\lambda_c, \lambda_\mp$ correspond to the points with error bar shown in Fig.~\ref{fig:transition}.}
\label{Table1}
\end{table}

\begin{figure}[htb]
\centering
	\includegraphics[width=0.9\linewidth]{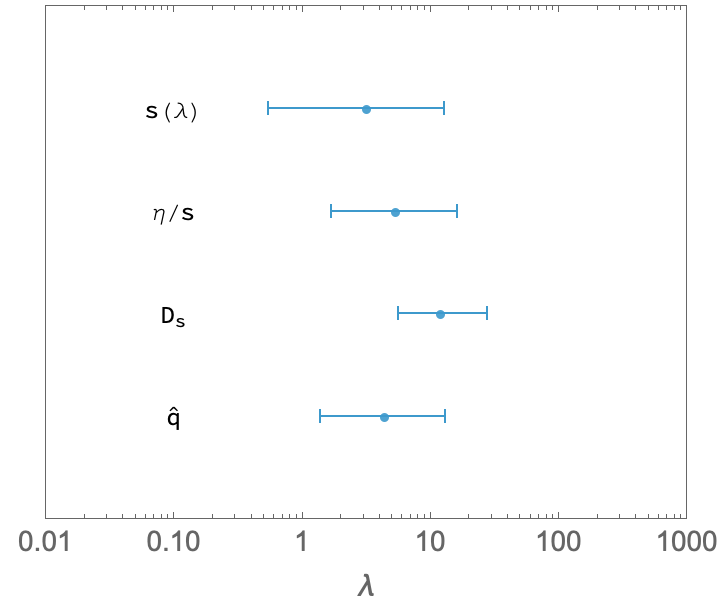}
\caption{Graphical representation of the central value $\lambda_c$ and range $\lambda_-\leq\lambda\leq\lambda_+$ of the weak-to-strong transition region for $s/s_0$, $\eta/s$, $2\pi TD_s$, and $\hat{q}/T^3$. The numercal values are listed in Table~\ref{Table1}.}
\label{fig:transition}
\end{figure}

Figure~\ref{fig:transition} graphically illustrates the central points and upper and lower bounds of the transition regions listed in Table~\ref{Table1}.

One can ask whether other definitions of $\lambda_c$ or different interpolation methods would have resulted in substantially different values. For example, in the cases of shear viscosity and the diffusion coefficient and $D_s$ we could have chosen the intersection point of the weak and strong coupling curves. For $\eta/s$ this would have given a value $\tilde\lambda_c = 3.01$ (instead of 9.51), and for $2\pi TD_s$ it would have resulted in $\tilde\lambda_c = 10.01$ (instead of 11.74). The rather large discrepancy in the case of $\eta/s$ may be due in part to the fact that we have used the scaling law as a substitution for the unknown perturbative NLO result. A different interpolation method could also have produced somewhat different valus for the central coupling and the boundaries of the transition regime. 

When trying to apply the insights from the ${\cal N}=4$ SYM theory to QCD, it is important to remember that the situation is somewhat different in QCD where the coupling runs as a function of scale. In thermal equilibrium this scale is set by the temperature $T$ and commonly evaluated at the lowest non-zero Matsubara frequency $\mu = 2\pi T$. For thermal coupling constants $\alpha_s(2\pi T) \approx 0.3$ the implied 't Hooft coupling $\lambda = 4\pi\alpha_s N_c \approx 11$ lies near the upper end of the transition region identified here. This makes the blanket use of the strong coupling limit of ${\cal N}=4$ SYM theory for modeling of bulk quark-gluon plasma physics questionable, even if finite coupling corrections are applied. Conversely, the use of thermal perturbation theory, even with the NLO HTL corrections, is likely to be also quantitatively unreliable.

In summary, we have constructed Pad\'e approximants between the weak and strong coupling regimes for the equation of state and three transport coefficients in ${\cal N}=4$ super-Yang-Mills (SYM) theory up to next-to-leading order in the 't Hooft coupling $\lambda$. In all four cases thransition is located in the region $3\leq\lambda\leq 14$. This range includes the value $\lambda\approx 12$ that has often been used in attempts to apply results from the SYM model to transport processes in the quark-gluon plasma. The conclusion from this study is that neither the strong coupling nor the weak coupling expansion is reliable in this physically most interesting range. 
\smallskip

{\it Acknowledgments:} I thank S. Waeber (Ben Gurion University) for helpful discussions and W.A. Zajc for useful comments. This work was supported by a grant from the U. S. Department of Energy, Office of Science (DE-FG02-05ER41367).

\subsection*{Appendix: Pad\'e approximants}

The generalized Pad\'e approximant for $s(\lambda)/s$ can be found in Appendix C od \cite{Andersen:2021bgw}. The approximants for the three normalized, dimensionless transport coefficients, which coincide with the weak and strong coupling expressions at NLO level when expanded to the same order in powers of the 't Hooft coupling, are:

\noindent (1) {\it Kinematic shear viscosity:}
\noindent The NLO weak coupling result is unknown; we use instead the expression $\eta/s = 2T^3/\hat{q}$ with the NLO result for $\hat{q}$, see eq.~(4.8) in \cite{Ghiglieri:2018ltw}. The NLO strong coupling result is given in eq.~(1.2) in \cite{Buchel:2008sh}. The generalized Pad\'e approximant is:
\begin{equation}
\frac{\eta(\lambda)}{s(\lambda)} = \frac{12\pi^2+aB\lambda +\lambda^2(A+B\sqrt\lambda)}{4\pi\lambda^2(A+B\sqrt\lambda)}
\end{equation}
with
\begin{eqnarray}
a &=& 15\zeta(3)
\nonumber \\
A &=& -3\ln(2\lambda)+7\frac{\zeta(3)}{\zeta(2)}\ln(q_{\rm max}/T)-0.4213 
\nonumber \\
B &=& 2.3539+\sqrt{2} 
\end{eqnarray}
where we choose $q_{\rm max}=10T$.

\noindent (2) {\it Diffusion constant:}
\noindent We consider the limit of a very slowly moving heavy quark and ignore velocity dependent terms that start at $O(v^2)$. The weak coupling NLO expression for the normalized diffusion constant can be gleaned from q.~(2.7) in \cite{Caron-Huot:2008dyw}:
\begin{equation}
\kappa_{\rm w}(\lambda) = \frac{\lambda^2T^3}{6\pi} \left( \ln\frac{1}{\sqrt{\lambda}} + 0.43044 + 0.80101 \sqrt{\lambda} \right).
\end{equation}
The strong coupling expression up to NLO was derived \cite{Casalderrey-Solana:2006fio}:
\begin{equation}
\kappa_{\rm s}(\lambda) = \pi\sqrt{\lambda}T^3 \left( 1- \frac{1}{2\lambda^{3/2}} + \ldots \right). 
\end{equation}
A good interpolation is obtained by setting
\begin{equation}
\kappa(\lambda) = \left( \frac{1}{\kappa_{\rm w}(\lambda)} + \frac{1}{\pi\sqrt{\lambda}} +\frac{1}{2\pi\lambda^2} \right)^{-1} .
\end{equation}

\noindent (3) {\it Momentum broadening parameter:}
\noindent For the NLO weak coupling expression for $\hat{q}/T^3$ see eq.~(4.8) in \cite{Ghiglieri:2018ltw}:
\begin{equation}
\hat{q}_{\rm w}(\lambda) = \frac{\lambda^2 T^3}{6\pi}\left( f(\lambda,q_{\rm max})+3.3289\sqrt{\lambda} \right)
\end{equation}
with 
\begin{equation}
f(\lambda,q_{\rm max}) = -3\ln(2\lambda)+7\frac{\zeta(3)}{\zeta(2)}\ln(q_{\rm max}/T)-0.4213 .
\end{equation}
Consistent with \cite{Ghiglieri:2018ltw} we choose $q_{\rm max} = 10T$ for the comparison.
The NLO strong coupling result can be found in eqs.~(1.2--4) in \cite{Zhang:2012jd}:
\begin{equation}
\hat{q}_{\rm s}(\lambda) = (\sqrt{\pi}T)^3 \frac{\Gamma(3/4)}{\Gamma(5/4)}(\sqrt{\lambda}-1.97) .
\end{equation} 
The Pad\'e approximant can be written as
\begin{equation}
\hat{q}(\lambda) = \frac{\hat{q}_{\rm w}(\lambda)}{1+\alpha\lambda^2+\beta\lambda^{3/2}},
\end{equation}
with
\begin{eqnarray}
\alpha &=& \frac{3.3289\,\Gamma(5/4)}{6\pi^{5/2}\Gamma(3/4)} ,
\nonumber \\
\beta &=& \frac{(f(\lambda,q_{\rm max})+6.5579)\,\Gamma(5/4)}{6\pi^{5/2}\Gamma(3/4)} .
\end{eqnarray}

\bibliographystyle{apsrev4-1}
\bibliography{Lessons}

\end{document}